\documentclass[fleqn,usenatbib]{mnras}

\usepackage{newtxtext,newtxmath}

\usepackage[T1]{fontenc}

\DeclareRobustCommand{\VAN}[3]{#2}
\let\VANthebibliography\thebibliography
\def\thebibliography{\DeclareRobustCommand{\VAN}[3]{##3}\VANthebibliography}

\usepackage{graphicx}	
\usepackage{amsmath}	

\usepackage{threeparttable}

\title[Dwarf galaxy candidates in nearby clusters]{The dwarf galaxy population in nearby clusters from the KIWICS survey }

\author[N. C. Choque-Challapa et al.]{
Nelvy Choque-Challapa,$^{1}$\thanks{E-mail: n.c.choque@astro.rug.nl}
J. Alfonso L. Aguerri,$^{2,3}$
Pavel E. Mancera Pi\~{n}a,$^{1,4}$
Reynier Peletier,$^{1}$
Aku Venhola,$^{5}$
\newauthor
and Marc Verheijen$^{1}$
\\
$^{1}$Kapteyn Astronomical Institute, University of Groningen, Landleven 12, NL-9747 AD Groningen, the Netherlands\\
$^{2}$Instituto de Astrof\'isica de Canarias, 38200 La Laguna, Tenerife, Spain\\
$^{3}$Universidad de La Laguna, Dept. Astrof\'isica, 38206 La Laguna, Tenerife, Spain\\
$^{4}$ASTRON, Netherlands Institute for Radio Astronomy, Postbus 2, 7900 AA Dwingeloo, the Netherlands\\
$^{5}$Space physics and astronomy research unit, University of Oulu, Pentti Kaiteran katu 1, 90014 Oulu, Finland
}
\date{Accepted XXX. Received YYY; in original form ZZZ}

\pubyear{2021}

\begin{document}
\label{firstpage}
\pagerange{\pageref{firstpage}--\pageref{lastpage}}
\maketitle

\begin{abstract}

We analyse a sample of twelve galaxy clusters, from the Kapteyn IAC WEAVE INT Cluster Survey (KIWICS) looking for dwarf galaxy candidates. By using photometric data in the $r$ and $g$ bands from the Wide Field Camera (WFC) at the  2.5-m Isaac Newton telescope (INT), we select  a sample of bright dwarf galaxies (M$_r$ $\leq$ -15.5 mag) in each cluster and analyse their spatial distribution, stellar colour, and as well as  their S\'ersic index and effective radius. We quantify the dwarf fraction inside the $R_{200}$ radius of each cluster, which ranges from $\sim$ 0.7 to $\sim$ 0.9. Additionally, when comparing the fraction  in the inner region with the outermost region of the clusters, we find that the fraction of dwarfs tends to increase going to the outer regions. We also study the clustercentric distance distribution of dwarf and giant galaxies (M$_r$ $<$ -19.0 mag), and in half of the clusters of our sample, the dwarfs are distributed in a statistically different way as the giants, with the giant galaxies being closer to the cluster centre. We analyse the stellar colour of the dwarf candidates and quantify the fraction of blue dwarfs inside  the   $R_{200}$ radius, which is found to be  less than $\sim$ 0.4, but  increases with  distance from the cluster centre.  Regarding the structural parameters, the  S\'ersic index for the dwarfs we visually classify as early type dwarfs tends to be higher in the inner region of the cluster. These results indicate the role that the cluster environment plays in shaping the observational properties of low-mass halos.

\end{abstract}




\section{Introduction}

Galaxy clusters are the largest and gravitationally bound structures in the Universe.  They cover a wide range of masses ranging from $\sim$ $10^{12}-10^{15}$ $M_{\odot}$, from small groups with tens of galaxies to clusters with thousands of members, although there is not a sharp line between them. In these high density environments, galaxies evolve under several physical processes that shape their morphology and stellar content. Therefore, galaxy clusters are natural laboratories to study galaxy evolution. 

The effect of the environment on galaxy evolution is reflected in different observational properties of field and cluster galaxies. For example, the morphology-density relation \citep{Dressler1980}, the different structural parameters observed in galaxies in clusters \citep{Aguerri2004}, or the so-called Butcher-Oemler effect \citep{ButcherOemler1978}. These differences between cluster and field galaxies can be explained as consequence  of several mechanisms acting in high density environments. Thus, strong tidal interactions between  galaxies or with the overall potential of the cluster/group can produce significant transformations among the cluster members \citep[e.g.][]{Moore1996, Gnedin2003, Smith2015}. In addition, the motion of the galaxies through the intracluster medium (or intragroup medium) can remove their cold and hot gas by different processes such as ram-pressure \citep{Quilis2000, Romanoliveira2019,Cortese2021}, starvation or strangulation \citep{Kawata2008, Fujita2004}.

Low-mass galaxies are the most abundant in the nearby Universe and within galaxy clusters \citep{Phillipps1998}.  They can be divided in two main groups: dwarf ellipticals (dE) and dwarf irregulars (dIrr). The former are characterised by regular shapes and red stellar colours \citep{Kormendy2009}, while dIrr show blue colours and irregular morphologies \citep[e.g.][]{vanZee2000}. Dwarf galaxies also follow the morphology-density relation. The number of dE increases with the galaxy density \citep{binggeli1988, Trentham2002}. In contrast, dIrr are mostly found in the outer cluster regions or in field environments \citep[e.g.][]{SanchezJanssen2008, Venhola2019}.

The formation and evolution of dwarf galaxies is still matter of debate, but it is well known that due to their shallow gravitational potential dwarf galaxies are, in general, more susceptible to the environment. Observational evidence suggests that dwarf ellipticals might be the result of galaxy transformations: gas-rich dwarf galaxies can be transformed into dE by the removal of their gas content  and the quenching of their star formation. In addition, bright galaxies can lose mass and be transformed into dwarf ones.  Several  external  process related with the environment  have been proposed to explain  these transformations against the  internal effects  \citep[see e.g., ][]{Boselli2006}.

In some nearby clusters a number of environmental transformation processes have been caught in the act. In the Hydra I cluster, \cite{Koch2012} reported an ongoing tidal disruption of a dwarf galaxy, indicating how tidal forces  can shape the morphological and kinematic properties of a dwarf galaxy. In a similar way, in the Virgo cluster, \cite{Kenney2014} reported a ram pressure stripping tail in the dIrr galaxy IC34188, a similar tail has been also reported in a dIrr  galaxy in  the  Cen A group  \citep{Johnson2015}.  

A number of observations support the scenario of the dwarf transformation from gas stripping \citep[e.g.][]{Janz2021}. For instance: $i)$ a similar faint-end slope of the spectroscopic galaxy luminosity function in clusters and field \citep[suggesting that the environment acts to modify the star formation activity;][]{Boselli2011, Agulli2014, Aguerri2020}; $ii)$ the relation between gas content, colour and cluster location \citep{Gavazzi2013}; $iii)$ the distribution in some nearby clusters of the dwarf post-starburst galaxies \citep{Aguerri2018}; $iv)$ the relation between age, metallicity and location in the clusters of dwarf galaxies \citep{Boselli2008, Smith2009, Toloba2009, Koleva2013}; $v)$ the presence of blue cores in some early-type dwarfs \citep{Lisker2007, Pak2014, Urich2017, Hamraz2019}. Nevertheless, the family of early-type dwarf galaxies is complex. Their formation might be influenced by different factors rather than a single effect \citep{Hamraz2019}. Some early-type dwarf galaxies show several morphological structures like disk features and blue centres \citep{Aguerri2005a,Lisker2007,Janz2014}, rotation \citep{Toloba2009}, or the presence of star formation, gas and dust in their centres \citep{Lisker2006a,Lisker2006b,Lisker2007}.

 Additionally, the cluster cores seem to be a hostile region for dwarfs \citep{SanchezJanssen2008,Pavel2018, Venhola2019}.  A number of studies have used  the dwarf-to-giant ratio (DGR) to quantify the fraction across the cluster, for example, \citet{Rude2020} combined  the luminosity functions (LFs) in the $r$ and $u$ bands of 15 Abell galaxy clusters and found that the dwarf-to-giant ratio increase at further clustercentric radius. Previously, \citet{Barkhouse2009} for  a sample of 57 low-redshift Abell clusters found a steady increase of the DGR with increasing clustercentric distance, however,  the differences in both works  may due to systematical definitions, as mentioned in \citet{Rude2020}. In any case, the change of the DGR  with the clustercentric distance could be due to the variation of giants, dwarfs or both as  pointed out in \citet{SanchezJanssen2008}. In addition, the DRG has been also studied as function of some cluster properties. Analysing 69 nearby clusters, \citet{Popesso2005} found a significant correlation  of the dwarf-giant ratio with the mass, velocity dispersion and X-ray luminosity, where the DGR tends to decrease as these  cluster parameters increase. This correlation becomes more significant by using a fixed radius  and is less significant when the $R_{200}$\footnote{$R_{200}$ defined as the radius within which the cluster density is 200 times the critical density of the Universe.} radius is used; similar results are also mentioned in \citet{Barkhouse2009}.

The evidences  described before have helped  us to understand more about  the transformation process and the various environmental effects  that dwarf galaxies can  experience along their evolution.  It is clear that studying  them in environments of different properties might also help to understand more about their evolution. In this work we use a sample of twelve nearby clusters with different properties. We study the effect of the environment on their galaxy population, particularly focusing on the dwarf galaxies regime. 

The structure of this paper is as follows. Section \ref{sec:sec2} contains the observation details of the survey and  a brief description of the data reduction process and  the cluster sample selection for this work. In Section \ref{sec:sec3}, we describe the galaxy sample selection. In Section \ref{sec:sec4}  and \ref{sec:sec5} the main results  and the discussion of them are presented, respectively. Finally, in Section \ref{sec:sec6} we summarise the main results of this work.

Along this work we use magnitudes in the AB system, and we adopt a $\Lambda$ cold dark matter ($\Lambda$CDM) cosmology  with $\Omega_m$=0.3,  $\Omega_{\lambda}$=0.7, and H$_0$= 70 km s$^{-1}$ Mpc$^{-1}$.

\section{Observations}
\label{sec:sec2}

The Kapteyn IAC WEAVE INT Cluster Survey (KIWICS, e.g. \citealt{Pavel2019}, PIs R. Peletier and A. Aguerri) is an observational campaign to obtain deep images of nearby galaxy clusters. It was carried out from 2016 until 2019 by using the Wide Field Camera (WFC) at the 2.5-m Isaac Newton Telescope (INT) at the Roque de los Muchachos observatory (ORM) in La Palma, Spain. The observational campaigns consisted in the imaging of 47 galaxy clusters selected from two X-ray flux limited catalogues compiled from the ROSAT All-Sky Survey dataset: the ROSAT Brightest Cluster Sample \citep[][BCS]{Ebeling1998} and its extension \citep[][eBCS]{Ebeling2000}. The BCS is 90\% complete for fluxes  higher than $4.4\times10^{-12}$ $\mathrm{erg}$ $\mathrm{cm^{-2}}$ $\mathrm{s^{-1}}$ in the ROSAT 0.1 - 2.4 keV band. The eBCS extends the BCS  down to $2.8\times10^{-12}$ $\mathrm{erg}$ $\mathrm{cm^{-2}}$ $\mathrm{s^{-1}}$ with 75\% of completeness. The full KIWICS sample encloses clusters from BCS and eBCS with redshifts between  0.01 and 0.04. These clusters were selected as targets of the future spectroscopic WEAVE Cluster Survey (WCS) that will be run at the 4.2-m William Herschel Telescope (WHT) at the ORM. This survey will be part of the astronomical surveys that the new WHT Enhanced Area Velocity Explores (WEAVE) spectrograph will carry out \citep{Dalton2016}.  The lower limit of the redshift range of the clusters was chosen based on the field-of-view (FOV = 2 deg diameter) of WEAVE,  while the upper-limit of the redshift range  was imposed by the WEAVE observational limiting magnitude ($m_{r} = 20.0$ mag; $M_{r} \approx -16.0$ mag at $z = 0.04$). Moreover, all the clusters are located in the Northern hemisphere, criterion imposed so that all clusters can be observed from the WHT with enough elevation in the sky.

\begin{figure}

	\includegraphics[width=\columnwidth]{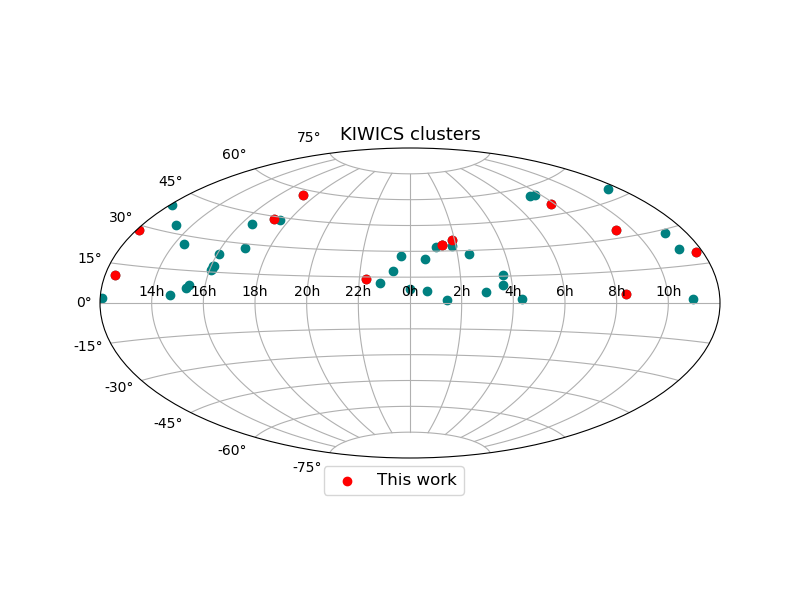}
    \caption{Sky Map of the full KIWICS sample. Red symbols correspond to the clusters analysed in this work.}
    \label{fig:figure1}
\end{figure}

KIWICS deep observations were carried out through the two broadband Sloan filters $r$ and $g$. Every cluster was observed with an integration time of  $\sim$ 5400 s and $\sim$ 1800 s in the $r$ and $g$ filters, respectively. The total integration time was split in exposures of 210 s each following a dithering pattern. The final coadded image covers at least an area of radius $R_{200}$ around the centres of each cluster.  Figure \ref{fig:figure1} and Table \ref{tab:table1} show the position in sky as well as the main properties for the full sample studied in this work, which correspond to the KIWICS clusters with redshift less than 0.029 (the remaining clusters of the survey will be studied in a future work). Further information  about the observations can be  found in \citet{Pavel2018, Pavel2019}.

\begin{table*}
	\centering
   \caption{Main properties of the clusters . Column 1  corresponds to their names, columns 2 and 3 shows their coordinates, column 4 their redshift and columns 5,6 and 7 to their mass ($M_{500}$), radius (R$_{500}$) and luminosity (L$_{500}$) values (from the \citealt{Piffaretti2011} catalogue).}
	\vspace{0.5cm}
	\label{tab:table1}
	\begin{threeparttable}
	\begin{tabular}{lcccccccr} %
		\hline
		Cluster name & RA (J2000) & DEC (J2000) & z & $M_{500}$ & R$_{500}$&L$_{500}$\tnote{a}\\
		        & (hh:mm:ss)    & ($^o$:':'')     & & $\times 10^{14} M_{\odot}$& Mpc&$\times10^{44}$ erg s$^{-1}$  \\
		\hline
		\hline

RXJ0123.2+3327 & 01:23:12.2 &33:27:40 & 0.0146 & 0.36   & 0.50 & 0.06 \\
A262           & 01:52:45.0 &36:09:25 & 0.0163 & 1.19   & 0.74 & 0.42 \\
RXJ0123.6+3315 & 01:23:41.0 &33:15:40 & 0.0164 & 0.61   & 0.60 & 0.14 \\
A1367           & 11:44:36.5 & 19:45:32 & 0.0214  & 2.14 & 0.90 &1.10\\
RXCJ0751.3+5012 & 07:51:22.5 & 50:12:45 & 0.0228 & 0.42& 0.52 &0.07\\
RXCJ0919.8+3345 & 09:19:49.2 & 33:45:37 & 0.0230 & 0.26 & 0.45 &0.08\\
RXCJ2214.8+1350 & 22:14:52.7 & 13:50:48 & 0.0253  & 0.32 & 0.48 & 0.05 \\
RXCJ1223.1+1037 & 12:23:6.50 & 10:37:26 & 0.0258  &0.56 & 0.58 & 0.12 \\
RXCJ1714.3+4341 & 17:14:18.6 & 43:41:23 & 0.0276 & 0.31 & 0.48 &0.05 \\
RXCJ1715.3+5724 & 17:15:21.9 & 57:24:27 & 0.0276 & 0.87 & 0.67 &0.25 \\
RXCJ1206.6+2811 & 12:06:37.4 & 28:11:01 & 0.0283 & 0.42&  0.52 &0.08\\
ZwCL1665        & 08:23:11.5 & 04:21:22 & 0.0293 & 0.73& 0.63 &0.19&\\
		\hline
			
	\end{tabular}
	\begin{tablenotes}
	    \item[a] On average the uncertainties on the L$_{500}$ measurements range from 15 to 20 per cent  \citep{Piffaretti2011}.
    \end{tablenotes}
    \end{threeparttable}
\end{table*}

\subsection{ Data reduction }

The reduction of the images was done with Astro-WISE \citep{Mcfarland2013} following the same procedure as explained in detail in \cite{Venhola2018} and in \cite{Pavel2019}. Here, we briefly summarise the main steps. 

The first phase of the data reduction consisted on the standard instrumental corrections; overscan,  bias, and flat-fielding corrections were applied in all the sciences frames.  The second phase took care of the sky subtraction followed by  the astrometry and photometric corrections. For this last part, standard star fields that were observed at every night of observation were used for the calibration. Finally, with the sciences images cleaned of bad pixels and cosmic rays, they were median stacked and coadded  into a single mosaic by using Swarp \citep{Bertin2010} and sampled to a pixel size of 0.2 arcsec.

\subsection{Sample selection and cluster properties}
In this work, we  analysed the clusters from the KIWICS sample with redshifts lower than 0.029. The poor quality of the final coadded image and the bad seeing of the observations made that seven clusters from the original sample were excluded, and that  twelve clusters from KIWICS sample were considered. The fourth and fifth columns of Table \ref{tab:table2} show the mean seeing for the $r$ and $g$ bands for each cluster. The twelve clusters considered in this study have a mean seeing of $\sim$ 1.4 arcsec in both $g$ and $r$ filters.

It also worth to mention that our final sample contains four clusters analysed already in  \cite{Pavel2019} where they studied eight clusters from KIWICS  searching for ultra-diffuse galaxies (UDGs).  In the present study  we will analyse the properties of the dwarf galaxies.  We  considered as dwarf galaxies those with $M_r >-19.0$ mag and giant galaxies those with $M_{r} < -19.0$ mag. This limit was used following the historical convention to classify dwarf galaxies as those with $M_{B} > -18.0$ mag \citep[][]{binggeli1988} and assuming a colour $B - r \sim 1.0$ mag for these galaxies.  We also considered a low luminosity cut at $M_{r} = -15.5$ mag. Galaxies fainter than this limit were not analysed to avoid strong contamination of background objects. 

\begin{figure}
\centering
	\includegraphics[width=\columnwidth, height=15cm]{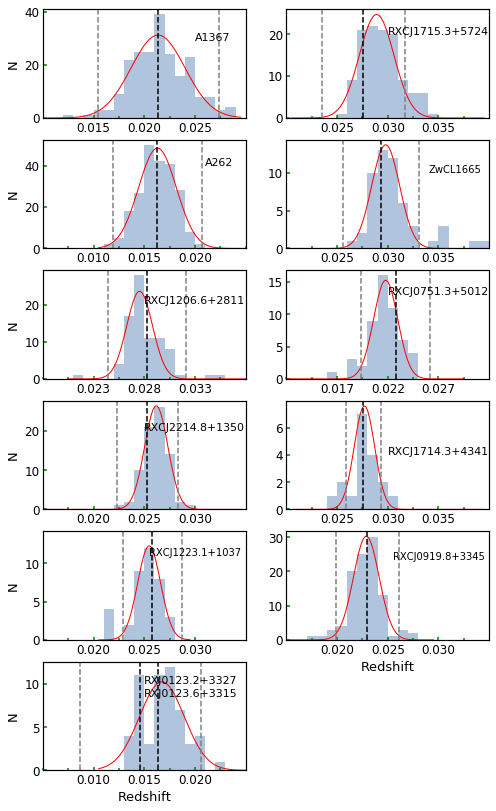}
    \caption{Redshift distributions for each cluster. The black dotted line marks the central redshift of the cluster, while the dotted grey lines indicate three times the velocity dispersion on each side. The fitted gaussian  for the objects between the grey dotted lines is also shown in red colour. Note that  the redshift distribution for clusters RXJ0123.2+3327 and RXJ0123.6+3315 are mixed (bottom panel) as they are in the same field of view. }
    \label{fig:figure2}
\end{figure}

Radial velocities for the galaxies in our clusters were  obtained from Sloan Digital Sky Survey (SDSS) , NASA/IPAC Extragalactic Database (NED), and data from WIYN telescope\footnote{Redshit data for cluster A262 comes from the WIYN 3.5-meter telescope, Hydra spectrograph. Private communication with J. Healy. }. This spectroscopic data was used to estimate the cluster membership, the mean velocity ($V_{c}$) and the velocity dispersion of each cluster ($\sigma_{c}$). Figure \ref{fig:figure2} shows the velocity histograms of the clusters and Table \ref{tab:table2} lists the velocity dispersion of the clusters. Both, the cluster membership and the velocity dispersion of the clusters were obtained by using a sigma clipping algorithm. In particular, we  used a 3-$\sigma$ clipping to determine the cluster membership and a 2-$\sigma$ clipping for the velocity dispersion. In Figure \ref{fig:figure2} it is highlighted  a  3-$\sigma_c$ range around the cluster central redshift (dashed vertical grey lines), which is the range we considered for cluster members. Note that there are  three clusters; RXCJ1714.3+4341, RXJ0123.2+3327 and RXJ0123.6+3315, the first one  with very few objects with spectroscopic redshifts available, and the last two  having a mixed redshift distribution  as they are in the same field of view.

\begin{figure}
\centering
	\includegraphics[width=\columnwidth, height=6cm]{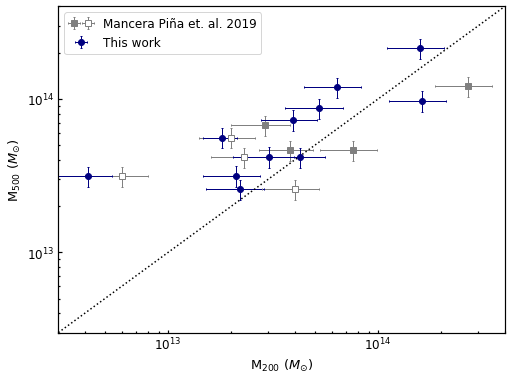}
    \caption{$M_{500}$ mass from \citet{Piffaretti2011} as a function of the $M_{200}$ mass we estimated from the velocity dispersion for each cluster. The filled and open grey points correspond to the clusters analysed in \citet{Pavel2019}, the open symbols corresponding to the four clusters that we have in common. The blue points correspond to the clusters analysed in this work. }
    \label{fig:figure3}
\end{figure}

The measured $\sigma_{c}$ was used to compute the $R_{200}$ radius of the clusters following $R_{200} = \sqrt{3} \sigma_v/10 H(z_c)$  \citep{Carlberg1997}. In addition, the halo mass ($M_{200}$) was obtained following the velocity dispersion-mass relation from \citet{Munari2013}. In general, the values of $\sigma_c$ are in a range  $\sim$ 175 km/s to $\sim$ 580 km/s. The clusters Abell 1367, RXCJ1715.3+5724, and Abell 262 have the largest values of the  velocity dispersion and are accordingly the most massive ones? (see Table \ref{tab:table1}). Figure \ref{fig:figure3} shows the $M_{200}$ mass we estimated for each clusters  as a function of their $M_{500}$ mass from \citealt{Piffaretti2011} (values from X-ray measurements). We also added the values of $M_{200}$ computed by \cite{Pavel2019} for their KIWICS subsample.  The uncertainties in $M_{200}$ were obtained by using Monte Carlo error propagation, while the uncertainty for  $M_{500}$  is assumed to be fifteen percent of the mass. This last assumption was done based on the average measurement uncertainties for L$_{500}$ explained in \citet{Piffaretti2011}, which range from 15 to 20 percent. Certainly, it is worth to mention that any under/sub estimation of the velocity dispersion may also affect the estimation  for $M_{200}$ and $R_{200}$ (e.g., clusters RXJ0123.2+3327 and RXJ0123.6+3315).

\begin{table*}

	\centering
	\caption{Properties of the clusters. Column 2 and 3 show the estimated values for the velocity dispersion and $M_{200}$. Columns 4 and 5 show the seeing of the coadd image in the $r$ and $g$ band. Last column shows the status classification we gave the cluster. The errors in the velocity dispersion values come from the Gaussian fit (Fig. \ref{fig:figure2}).}
	\vspace{0.5cm}
	\label{tab:table2}
	\noindent
	\begin{threeparttable}
	\begin{tabular}{lcccccr} %
		\hline
		Cluster & velocity dispersion & $M_{200}$& seeing $r$-band & seeing $g$-band &status\\
		&  (km/s)  &  $\times 10^{13} M_{\odot}$& (arsec) & (arsec) &(relaxed?)  \\
		       
		\hline
		\hline
RXJ0123.2+3327 & 483 $\pm$95  & 0.5 $\pm$ 0.2\tnote{a}  & 1.6& 1.3 & u\tnote{b} \\
A262           & 402 $\pm$24  & 5.0 $\pm$ 1.5           & 1.5 & 1.4 &u&\\
RXJ0123.6+3315 & 483 $\pm$95 & 0.5 $\pm$ 0.2\tnote{a}  &1.6 & 1.3&u\tnote{b} \\
A1367           &  581 $\pm$64 & 15.0 $\pm$ 4.6        &1.5  &1.6&n  \\
RXCJ0751.3+5012 & 360 $\pm$24 & 3.5 $\pm$ 1.1          &1.3 & 1.3&n \\
RXCJ0919.8+3345 & 299 $\pm$19  & 2.0 $\pm$ 0.6         &1.4 &1.4 &y\\
RXCJ2214.8+1350 & 351 $\pm$07 & 3.3 $\pm$ 1.0           & 1.3& 1.3&u \\
RXCJ1223.1+1037 &  302 $\pm$12 & 2.1 $\pm$ 0.6         &1.6 & 1.5 &y\\
RXCJ1714.3+4341 & 176 $\pm$22 & 2.6 $\pm$ 0.8          &1.3 &1.5&y\\
RXCJ1715.3+5724 & 475 $\pm$29 & 8.1 $\pm$ 2.6          & 1.3 &1.5 &u \\
RXCJ1206.6+2811 &  381 $\pm$42 & 4.2 $\pm$ 1.3         &  1.5& 1.4&n \\
ZwCL1665        & 382 $\pm$15 & 4.3 $\pm$ 1.3          & 1.3& 1.4 &u\\

		\hline
	\end{tabular}
	
	\begin{tablenotes}
	    \item[a] Same estimated mass for RXJ0123.2+3327 and RXJ0123.6+3315.
        \item[b] Status unknown.
        
    \end{tablenotes}
    \end{threeparttable}
\end{table*}

To complement  the cluster properties we also looked at their dynamical status, whether it might be a relaxed system or not, to later see if there is any correlation with their dwarf properties. To identify reliably relaxed and unrelaxed clusters different approaches are commonly employed. The relaxation state is often inferred from the X-ray morphology \citep[e.g.][]{Mantz2015} and the distribution of relative velocities in clusters \citep[e.g.][]{Ribeiro2013}. Here,  we performed a visual check on  the X-ray contour maps available in the literature \citep{Kim1995,Jones1999,Dahlem2000,Rusell2007,Eckmiller2011,Russell2014} and classified the clusters as relaxed when the X-ray contour have circular shape. We used the Shapiro test to check whether the galaxy velocity distribution of each cluster has a normal distribution (see the result in Table \ref{tab:table3} and Fig. \ref{fig:figure2}). We classified a cluster (last column on Table \ref{tab:table2}) as relaxed if it satisfies both tests, otherwise we just classified it as unknown status.

\begin{table}
	\centering
   \caption{ Results of the Shapiro  test to check  the normality of the redshift distribution for the galaxies inside 3-$\sigma_c$ (Fig. \ref{fig:figure2}).  A p-value  less than significance level (often 0.05) means  we can reject null hypothesis that it is normally distributed.}
	\vspace{0.5cm}
	\label{tab:table3}
	\begin{tabular}{lcr} 
		\hline
		Cluster name & p-value  \\
		\hline
		\hline
RXJ0123.2+3327 & 0.17 \\
A262 &  0.27\\
RXJ0123.6+3315 &0.17 \\
A1367 & 0.04 \\
RXCJ0751.3+5012 & 0.05 \\
RXCJ0919.8+3345 & 0.41 \\
RXCJ2214.8+1350 & 0.22 \\
RXCJ1223.1+1037 & 0.72 \\
RXCJ1714.3+4341 & 0.94 \\
RXCJ1715.3+5724 &0.03  \\
RXCJ1206.6+2811 &0.01 \\
ZwCL1665 & 0.62\\

		\hline
			
	\end{tabular}
\end{table}

\section{SExtractor catalogues}
\label{sec:sec3}
The detection and the measurements of the photometric properties of the objects in the scientific images was done by using Source Extractor   \citep[SExtractor;][]{Bertin1996} in dual mode: the $r$-band images of the clusters were used for the detection and the measurements of the photometric parameters of the objects in this band. The $g$-band parameters were measured at the position of the objects detected in the $r$-band image. In addition, SExtractor was run using two different configurations. In the first pass, the configuration  used was similar to the one  by \cite{Pavel2019}. This configuration is optimised to detect small and faint galaxies. In the second pass, we used a different configuration to detect large and bright galaxies (Appendix \ref{sec:appA}). The final catalogues were a combination of the two SExtractor runs. For the common  objects in both catalogues, their properties were taken from the faint catalogue.

The criteria to select  galaxies from the full detection catalogues was made based on the SExtractor  stellarity $\mathrm{CLASS\_STAR}$\footnote{This parameter go from 0 to 1, objects close to 1 are more likely to be point sources and objects close to 0 are those who are more likely to be extended objects.} and FLAG parameters. We consider as galaxies those objects with CLASS$\_$STAR $\leq$ 0.2 in both filters $g$ and $r$ and with $\mathrm{FLAGS_{r} = 0}$,  that ensure that we are not considering objects that are blended with any other close object or without accurate photometry.

The zeropoint calibration was applied to the MAG$\_$AUTO magnitude. It was obtained using the stars (objects with $\mathrm{CLASS\_STAR_{g,r} \geq 0.8}$ and $\mathrm{FLAGS_r =0}$) located in our fields and with magnitudes measured in SDSS. Only one cluster, A262, was not in the SDSS footprint. In this case, we used the Pan-STARRS survey \citep{Chambers2016a} to calibrate the photometry. The final calibration contains also the Galactic extinction correction. This was obtained by using the  extinction calculator of NED \citep{Schlafly2011}. We did not apply k-correction to our magnitudes since its effect is not significant at the redshifts of our clusters (see for example \citealt{Chilingarian2010}).

\subsection{Background decontamination}
\begin{figure}
    \centering
\includegraphics[width=\columnwidth, height=10cm]{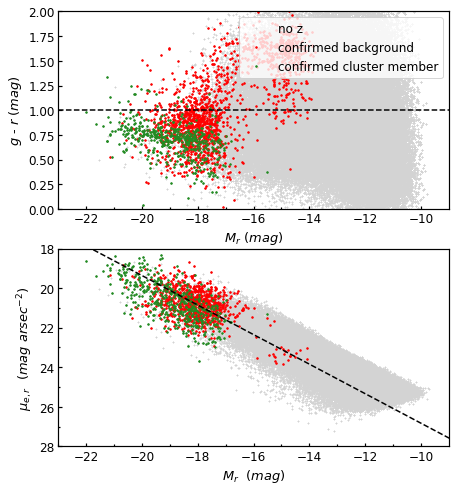}
\caption{Top panel: colour-magnitude diagram for the galaxies of all the clusters grey symbols. The  black dotted  line highlights the colour $g-r$ $\leq$ 1.0 mag used to discard background galaxies.  
Bottom panel: Mean surface brightness of the galaxies that  have passed the first colour cut as a function of their absolute magnitude (grey symbols).  For a fixed absolute magnitude, we considered cluster members to be those galaxies below the black dotted line. See text for more details. In both panels, green and red points represent spectroscopically confirmed cluster members and background galaxies, respectively.}
\label{fig:figure4}
\end{figure}

In order to remove background objects of the SExtractor catalogue, firstly we imposed a cut in colour $g - r$ $\leq$ 1.0 mag excluding all galaxies redder than this colour limit, which correspond to a 12 Gyr old stellar population with [Fe/H] = +0.25 supersolar metallicity \citep{Worthey1994}. The fraction of cluster members with $g - r > 1.0$  mag is expected to be small according to the galaxy colour distribution in the nearby Universe \citep{Hogg2004, Rines2008, Venhola2018}.  Figure \ref{fig:figure4}  shows the colour-magnitude diagram (CMD) for the galaxies of all clusters of the sample in the whole field of view of the images and the cut in colour we imposed (dashed line in the top panel).

However, we might still have some contamination of foreground and background galaxies  even applying a colour cut, as they can have a $g$ - $r$ $<$ 1.0 mag so we performed a second cut using the surface brightness-magnitude plane (similar to \citealt{Venhola2018}). To perform this selection, all the  clusters of our sample were analysed together (bottom panel; Fig.\ref{fig:figure4} ) in order to have a robust number of spectroscopically confirmed  (and not confirmed) galaxy members. We fitted a linear relation of $M_{r} - \mu_{e}$ described by the galaxy cluster members (green points in Fig. \ref{fig:figure4}). We considered as background objects those galaxies located  more than 1.5-$\sigma$ (black dashed line in Fig. \ref{fig:figure4}) above the linear relation in the $M_{r} - \mu_{e}$ defined by the cluster members.

\subsection{Visual classification}

We also implemented a visual classification of the galaxies in our sample brighter than $m_{r} = 20$ mag. Objects fainter than this magnitude are difficult to classify. The visual classification was also used to remove background galaxies that were not detected by using the colour and surface brightness cuts described previously. Following similar visual classifications in clusters \citep[see][]{Venhola2019, Wittmann2020}, we visually classify the galaxies in five groups: late type, early type, background, interacting and unknown.  Our classification is as follows,

\begin{itemize}

  \item Early type galaxies:  Galaxies with a uniform red colour (no blue patches) and morphology, not showing any clear spiral structure.
  
    \item Late type galaxies. In this category were included blue galaxies, showing structure features. In particular, the fainter objects that were included in this group were those that did not have a uniform colour, that is, with some characteristics such as blue spots.

  \item Likely background objects: Any object showing an irregular shape (with possible features like bars or spiral  arms) was included in this category.

    \item Unknown. Any object that is an artefact or an star. Also objects that due to their faint magnitude are difficult to include in any category.
    
    \item Likely interacting objects. Any object that seems to be merging or interacting with another one.
\end{itemize}

The final sample  excludes all the objects we classified as background,  artefacts, interactions or unknown objects and, as mention, all objects with  $m_r$ $>$ 20 mag, as the visual classifications there become not accurate. In Table \ref{tab:table4} we provide an overview of the number of the final sample down to an absolute magnitude\footnote{Converted from m$_r$ assuming that all the galaxies are at the mean distance of their associated clusters.}, M$_r$ $\leq$-15.5 mag.

\begin{table}
 \centering
   \caption{Number of  the final sample (dwarf plus giant galaxies) down to an absolute magnitude, M$_r$ $\leq$ -15.5 mag.}
	\vspace{0.5cm}
	\label{tab:table4}
	\begin{tabular}{lcr} 
		\hline
		Cluster & final sample   \\
		        & (after visual classification) \\
		\hline
A1367            &  542 \\
RXCJ1715.3+5724  &  306  \\
A262             &  468  \\
ZwCL1665         &  290  \\
RXCJ1206.6+2811  &  446 \\
RXCJ0751.3+5012  &  166 \\
RXCJ2214.8+1350  &  175   \\
RXCJ1714.3+4341  &  304  \\
RXCJ1223.1+1037  &  217  \\
RXCJ0919.8+3345  &  343   \\
RXJ0123.2+3327  &  248  \\
		\hline
			
	\end{tabular}
\end{table}

\subsubsection{Background contamination}
\begin{figure}
	\includegraphics[width=\columnwidth]{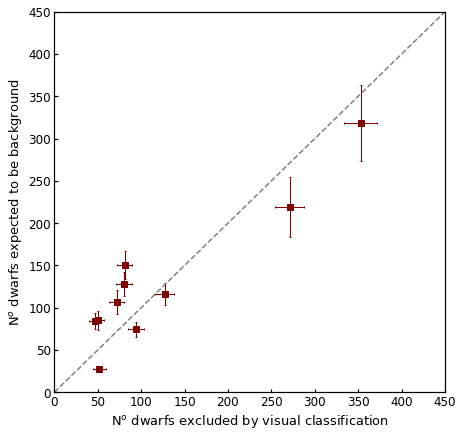}%
    \caption{Number of dwarfs excluded by visual classification as function of the number of dwarfs expected to be background galaxies (from a blank field). Error bars correspond to Poisson uncertainties based on the actual number of objects. }
    \label{fig:figure5}
\end{figure}

In order to quantify how much contamination may still be present in the final sample, particularly in the dwarf regime (-19.0 mag $\leq$ M$_r$ $\leq$ -15.5 mag) despite the visual classification we did, we estimated statistically the number of background galaxies  we might expect from a blank field. The field used for this purpose\footnote{CaBlank1 (01:47:36, +02:20:03). WFC blank fields catalogue; http://www.ing.iac.es/astronomy/instruments/wfc/blanks.html} was observed in the same way as in the clusters and  we used the same criterion as in the clusters to select galaxies from its SExtractor catalogue, this is same colour and surface brightness cut. We selected a circular area of 0.38 deg$^2$ by using the image in the $r$-band, and estimated the number of dwarf galaxies  present in the field and later we re-scaled this number to the area of each cluster. In Figure \ref{fig:figure5} we show the number of dwarfs expected to be background as a function of the number of dwarfs that we excluded in our visual classification.  The figure shows that the number of dwarfs removed using visual classification agrees reasonably well with the one expected from the blank field.

\section{Results}
\label{sec:sec4}

\subsection{Distribution of dwarf and giant  galaxies in clusters}

\begin{figure*}
    \centering
\includegraphics[width=15cm, height=20cm]{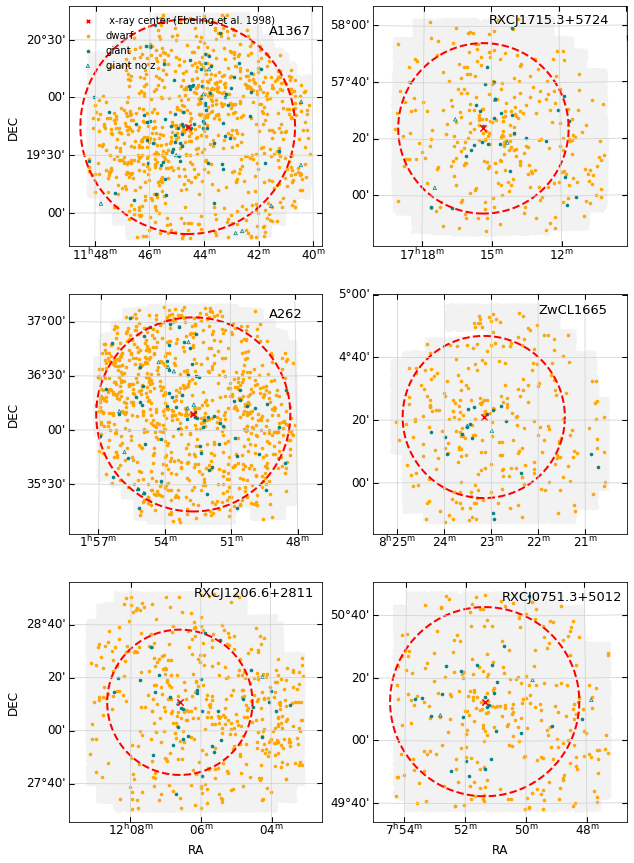}
\caption{ Distribution of dwarfs (yellow) and giants (teal) galaxies for each cluster. Teal circles correspond to giants for which redshift are available (mostly from SDSS) while teal open triangles correspond to giants without redshift available to confirm their cluster membership. A red marker highlights the X-ray centre from \citet{Ebeling1998}, the red circle indicates the estimated $R_{200}$ radius, and the grey area indicates the field covered by the image in the $r$-band. (continued). }
\label{fig:figure6}
\end{figure*}

\renewcommand{\thefigure}{\arabic{figure} (Cont.)}
\addtocounter{figure}{-1}

\begin{figure*}
    \centering
\includegraphics[width=15cm, height=20cm]{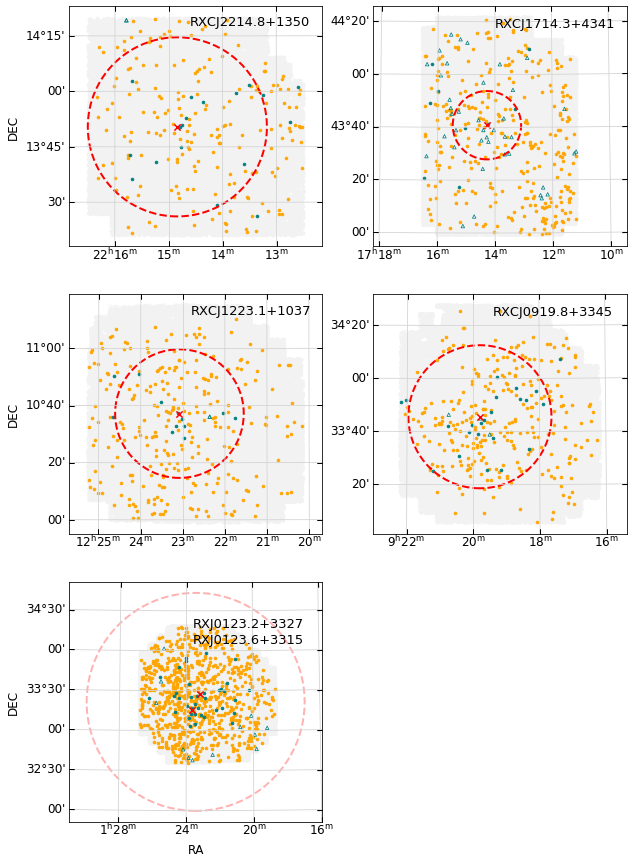}
\caption{ Distribution of dwarfs (yellow) and giants (teal) galaxies for each cluster. A red  marker highlights the X-ray centre from \citet{Ebeling1998}, the red circle indicates the estimated $R_{200}$ radius, and the grey area indicates the field covered by the image in the $r$-band. Note that in the last panel the two red markers correspond to two different X-ray defined clusters of our sample that are in the same field of view.}
\label{fig:figure6con}
\end{figure*}

 There are links between galaxy properties and their location in nearby galaxy clusters; for instance the morphological segregation observed in galaxy aggregations like Virgo \citep{Lisker2007}, Coma \citep{Aguerri2004}, Fornax \citep{Venhola2019} or other nearby clusters \citep{SanchezJanssen2008}. This segregation is observed in a wide range of galaxy luminosities.

 In Figure \ref{fig:figure6} and \ref{fig:figure6con}  we show the  spatial distribution of the dwarf  ($M_r$ $\geq$ -19.0 mag, orange points) and giant ($M_r$ $<$ -19.0 mag, green points) galaxies in the  field of view of each cluster.  Note that,  all the clusters are covered at least until 1$R_{200}$, where it is  also highlighted the X-ray centre (red cross symbol)  for each cluster \citep{Ebeling1998,Piffaretti2011}.

A visual inspection of  Figs. \ref{fig:figure6} and \ref{fig:figure6con} shows that dwarf galaxies tend to be widely distributed across the clusters but not always homogeneously, as in some cases they tend to  cluster in certain specific regions. For example, in A1367, which is the most massive and populated cluster in our sample, dwarfs tend to be localised in the central region but also in the substructures identified in this cluster the NE and SW regions \citep[e.g.][]{Ge2020}. Giant galaxies in this cluster follow the same distribution. In the cluster RXCJ1715.3+5724 dwarfs tend to be more concentrated in the central region, a similar distribution is also seen in clusters ZwCL1665, RXCJ1206.6+2811, RXCJ1714.3+4341, RXCJ0751.3+5012, RXCJ1223.1+1037, RXCJ2214.8+1350, RXCJ1714.3+4341 and  RXCJ0919.8+3345.  A262 shows a wide distribution of their dwarf galaxies  across the cluster but with some over densities in the border of the $R_{200}$ radius that might be due to the fact that this cluster is embedded in the Perseus supercluster so there  is large substructure of galaxies around it. With respect to the clusters RXJ0123.2+3327 and RXJ0123.6+3315, since both are in the same field of view, their population of galaxies is mixed and because there are not enough redshift for the galaxies in those clusters, it is not easy to analyse their dwarf and giant distributions separately.

We used a Kolmogorov-Smirnov (KS) test to analyse the statistical  differences between the giant and dwarf distributions within  $R_{200}$. Figure \ref{fig:figure7} shows the cumulative distribution function of the clustercentric distance for both giant and dwarf distributions on each cluster.  Table \ref{tab:table5} contains the values of the probability given by the KS test. In particular, five clusters (A1367, RXCJ1715.3+5724 , A262, ZwCL1665 and RXCJ1223.1+1037) show dwarf and giant radial distributions which are statistically different. In these cases, the bright galaxies are located closer to the cluster centre.  Moreover, these cases correspond to the more massive ones in our sample (as given by $M_{500}$ mass); other parameters such as velocity dispersion ($M_{200}$ indicator)  or the dynamical status do not show any clear difference. The remaining five clusters (RXCJ1206.6+2811, RXCJ0751.3+5012, RXCJ2214.8+1350, RXCJ1714.3+4341, and  RXCJ0919.8+3345) show a statistically similar radial distribution of dwarf and giant galaxies. The clusters RXJ0123.2+3327 and RXJ0123.2+3327  were not included in this figure (and in the following figures) as their field of view overlap and $R_{200}$ is not well determined. 

Note that in some clusters (i.e. ZwCL1665, RXCJ1206.6+2811, RXCJ0751.3+5012, RXCJ2214.8+1350, RXCJ1714.3+4341, and RXCJ0919.8+3345) the distributions of the giants start at larger distances, not necessarily at the cluster centre. The reason for this is an small offset between the X-ray centre we are using \citep[][;ROSAT observations]{Ebeling1998,Ebeling2000} and the position of of the brightest cluster galaxy. 

\renewcommand{\thefigure}{\arabic{figure}}
\begin{figure}

	\includegraphics[width=\columnwidth, height=16cm]{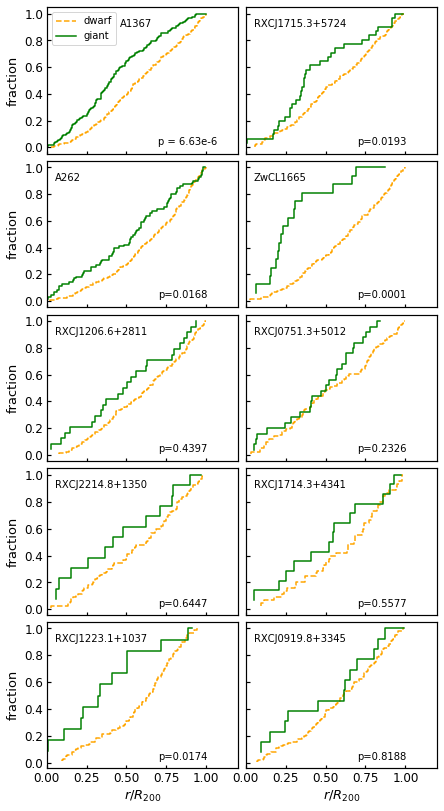}
    \caption{ Individual cumulative functions of the clustercentric distances of dwarfs (yellow dashed line) and giants (green line) on each cluster. The number at the corner of each panel corresponds to  the p-value  coming from a Kolmogorov-Smirnov test  which gives the probability that giants and dwarfs are taken from the same radial distribution.}
    \label{fig:figure7}
\end{figure}

\subsection{The dwarf galaxy fraction}
We  quantified the dwarf galaxy fraction (defined as the ratio between the number of dwarfs over  the total number of galaxies - dwarfs plus giants - within $R_{200}$) for each cluster as shown in Table \ref{tab:table5} (column 1). The dwarf fraction ranges from $\sim $0.75. to $\sim$ 0.91  in all clusters, with small scatter at a fixed mass. Apparently, there is not relation between the dwarf fraction and the dynamical status, mass or velocity dispersion of the clusters (see Table \ref{tab:table2}). We  discuss this further in Section 5.

\begin{table*}
	\centering
   \caption{Dwarf fraction and blue dwarf fraction on each cluster. Column 2 corresponds to the dwarf fraction, while column 3 corresponds  to the p-value resulting  after applying  a Kolmogorov-Smirnov test for clustercentric distance distributions between dwarf and giants (Fig. \ref{fig:figure7}). Column 4 corresponds to the blue dwarf fraction (Fig. \ref{fig:figure12}), while column 5 corresponds to the p-value after applying a KS test between the clustercentric distribution of blue and red dwarfs. Errors correspond to a binomial uncertainties.}
	\vspace{0.5cm}
	\label{tab:table5}
	\begin{tabular}{lcccr} 
		\hline
		Cluster & dwarf fraction  &giant-dwarf galaxies& blue dwarf fraction &blue-red dwarfs \\
		        & &(p-value)&  &(p-value)   \\
		\hline
A1367            &  0.75 $\pm$0.02& 6.6e-6 &0.29 $\pm$0.02& 0.0063  \\
RXCJ1715.3+5724 &   0.86 $\pm$0.02& 0.0193 &0.10 $\pm$0.02& 0.0761  \\
A262             &  0.82 $\pm$0.02& 0.0168 &0.21 $\pm$0.02& 0.3974 \\
ZwCL1665         &  0.91 $\pm$0.02& 0.0001 &0.31 $\pm$0.04& 0.0542  \\
RXCJ1206.6+2811 &   0.88 $\pm$0.02& 0.6584 &0.17 $\pm$0.03& 0.1811    \\
RXCJ0751.3+5012 &   0.79 $\pm$0.04& 0.7223 &0.24 $\pm$0.04& 0.8750   \\
RXCJ2214.8+1350 &   0.87 $\pm$0.03& 0.6447 &0.18 $\pm$0.04& 0.1208    \\
RXCJ1714.3+4341 &   0.76 $\pm$0.06& 0.5777 &0.16 $\pm$0.05& 0.8012  \\
RXCJ1223.1+1037 &   0.89 $\pm$0.03& 0.0174 &0.16 $\pm$0.04& 0.4329  \\
RXCJ0919.8+3345 &   0.85 $\pm$0.03& 0.8188 &0.31 $\pm$0.04& 0.0003  \\
		\hline
			
	\end{tabular}
\end{table*}

\subsection{Structural parameters for dwarf and giants }

The dependence of the distribution structural parameters on the environment has been studied in some nearby galaxies clusters, such as the Coma cluster \citep[see][]{Trujillo2004, Venhola2019,Janz2014}. We used {\small{GALFIT}} \citep{Peng2010} to determine the structural parameters of the galaxies in our clusters. In particular, we run {\small{GALFIT}} as is described in \cite{Venhola2018} for all the  possible cluster members  but excluding all the objects with the  SExtractor parameter ISORAREA-IMAGE being smaller than 200 pixels. This cut was mainly done to exclude smaller  objects  that are likely to have a bad fit as they are very faint (apparent $r$-band magnitudes fainter than $\sim 22$ mag). All the objects were fitted using a single S\'ersic function in both $r$ and $g$ bands. Given the resolution of our images we were not able to distinguish a nucleus in the galaxies.  We inspected the best fits by checking the output parameters, and considered as bad fits those with  effective radius close to zero, also were excluded those objects with a  significant (more than 0.5 mag) difference between SExtractor and GALFIT magnitudes. 

\begin{figure}

	\includegraphics[width=\columnwidth, height=17cm]{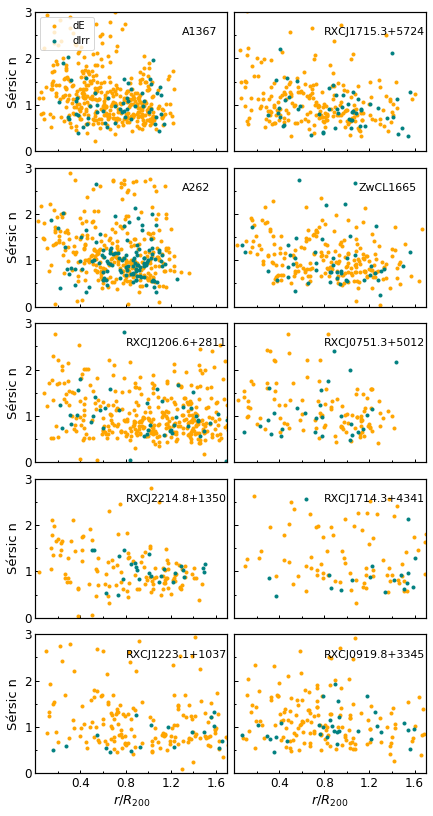}
    \caption{S\'ersic index ($n$) as a function of clustercentric distance for early type dwarfs (orange points) and late type dwarfs (green points). Note that cluster RXJ0123.2+3327 and RXJ0123.2+3315 are not included in this figure.} 
    \label{fig:figure8}
\end{figure}

Figure  \ref{fig:figure8} shows the S\'ersic index ($n$) of the dwarf galaxies as a function of the clustercentric distance. We divided the dwarf sample according to the visual classification (described in Section 3.2), in early (dE) and late type dwarfs (dIrr). In six (A1367, RXCJ1715.3+5724, A262, ZwCL1665, RXCJ1206.6+2811, RXCJ1223.1+1037) out of the ten clusters there seems to be a trend  in the values of  the S\'ersic index for the dE galaxies. In particular, the values of $n$ increase towards the inner regions of the clusters as  shown in Table \ref{tab:table6}, which lists the mean value of $n$ computed in the inner ($r$ $\leq$ 0.3$R_{200}$)\footnote{We split the sample at this clustercentric distance to better contrast the population in the inner  and outer regions of the cluster.  } and the outer cluster regions (0.3$R_{200}$ $<$ $r$ $\leq$ $1R_{200}$). This indicates that dwarfs located in the inner cluster regions have more concentrated light profiles than those located in the outer regions. This result can be explained if galaxies in the inner cluster regions are more affected by environmental effects than those located in the outskirts of the clusters \citep[e.g.][]{Trujillo2002}.  It is also possible that more concentrated dwarfs have survived the cluster environment of the inner regions while those with a flatter profile have not. The trend observed for the values of $n$ in the dE sample is not present for the late-type dwarf galaxies.

\begin{table*}
	\centering
   \caption{Mean value for the S\'ersic index $n$  (columns 2 and 3) and the effective radius (columns 4 and 5)  measured in bins of clustercentric distance, $r/R_{200}\leq$0.3 and 0.3$<r/R_{200}\leq$1.0. The error value corresponds to the standard deviation measured in each distance bin normalised by the square root of the total value on each bin. Note that these values are for those dwarfs classified as dwarf ellipticals. }
	\vspace{0.5cm}
	\label{tab:table6}
	\begin{tabular}{lcccr}
		\hline
		Cluster & mean $n$&  mean $n$ &mean $R_{\rm eff}$ (kpc)& mean $R_{\rm eff}$ (kpc)\\
		  & $r/R_{200}\leq$0.3  &  0.3$<r/R_{200}\leq$1.0 &$r/R_{200}\leq$0.3 &0.3$<r/R_{200}\leq$1.0 \\ 
		
		\hline
A1367            & 1.52 $\pm$ 0.09 & 1.21 $\pm$ 0.04  & 1.32$\pm$  0.07 & 1.30$\pm$  0.03 \\
RXCJ1715.3+5724 & 1.38 $\pm$ 0.11 & 1.09$\pm$  0.04  & 1.49$\pm$  0.08 & 1.32$\pm$  0.05 \\
A262            & 1.54$\pm$  0.09 & 1.26$\pm$  0.05 & 1.23$\pm$  0.09 & 0.90$\pm$  0.03 \\ 
ZwCL1665       & 1.50$\pm$  0.08 & 1.14$\pm$  0.08  & 1.38$\pm$  0.10 & 1.24$\pm$  0.04 \\   
RXCJ1206.6+2811 & 1.53$\pm$  0.12 & 1.08$\pm$  0.06  & 1.26$\pm$  0.07 & 1.26$\pm$  0.04 \\
RXCJ0751.3+5012 & 1.46$\pm$  0.15 & 1.29$\pm$  0.09  & 1.42$\pm$  0.10 & 1.20$\pm$  0.07 \\
RXCJ2214.8+1350 & 1.38$\pm$  0.09 & 1.12$\pm$  0.08 & 1.27$\pm$  0.08 & 1.28$\pm$  0.07 \\
RXCJ1714.3+4341 & 1.56$\pm$  0.15 & 1.32$\pm$  0.07 & 1.46$\pm$  0.12 & 1.32$\pm$  0.06 \\
RXCJ1223.1+1037 & 1.56 $\pm$  0.19 & 1.36$\pm$  0.27 & 1.39$\pm$  0.16 & 1.30$\pm$  0.16 \\
RXCJ0919.8+3345 & 1.27$\pm$  0.13 & 1.30$\pm$  0.07  & 1.29$\pm$  0.09 & 1.37$\pm$  0.07 \\
		\hline
			
	\end{tabular}
\end{table*}

\begin{figure}

	\includegraphics[width=\columnwidth, height=16.4cm]{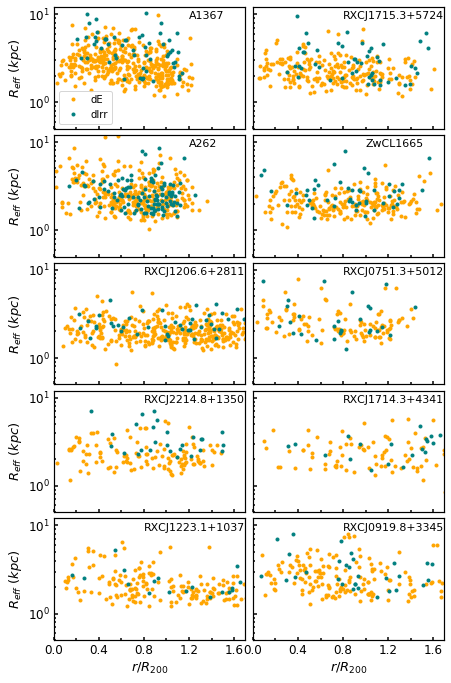}
    \caption{Effective radius ($R_{\rm eff}$) as a function of clustercentric distance for early type dwarfs (orange points) and late type dwarfs (green points). Note that cluster RXJ0123.2+3327 and RXJ0123.2+3315 are not included in this figure.}
    \label{fig:figure9}
\end{figure}

In addition to the S\'ersic index, in Figure \ref{fig:figure9} we also show the effective radius ($R_{\rm eff}$) as a function of the  clustercentric distance, and similarly we quantify the mean value of $R_{\rm eff}$ in the inner and the  in outermost region of the clusters (see Table \ref{tab:table6}). Although most of the clusters show a slightly higher mean value of $R_{\rm eff}$ in the inner regions, these differences are smaller than 2-$\sigma$.

 \subsection{The $g-r$ colour of galaxies}

\begin{figure}
   \centering
	\includegraphics[width=\columnwidth, height=6cm]{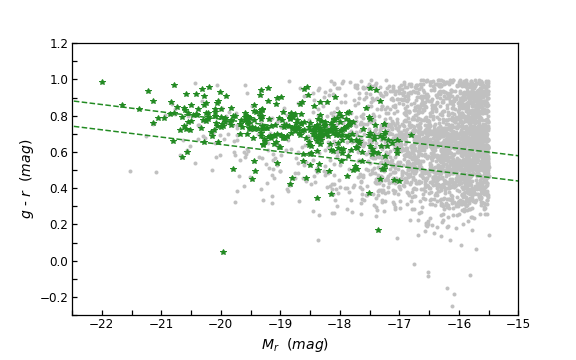}
    \caption{Colour-magnitude diagram for all the red sequence elliptical galaxies (green points) of all cluster of the sample espectrally confirmed as a cluster member. Upper green line corresponds to  the fit line for these galaxies while the bottom green  line is the same line shifted three  sigmas (standard deviation of the distance of the the green points to the fitting line) below. Grey points correspond to  the remaining galaxies from each cluster.}
    \label{fig:figure10}
\end{figure}

We classified the galaxies in blue and red according to their $g - r$ stellar colour. This classification was done according to their position in the CMD. Figure \ref{fig:figure10} shows the CMD of the galaxies in all clusters. The green points represent those galaxies confirmed spectroscopically as cluster members. Using an iterative sigma-clipping algorithm, we fitted the red sequence of this global CMD that was used to split between blue and red galaxies. We considered as blue galaxies those with a $g - r$ colour bluer than 3-$\sigma$ (bottom line), being $\sigma$ the dispersion in colour of the galaxies around the red sequence.

\begin{figure*}

	\includegraphics[width=17cm, height=15cm]{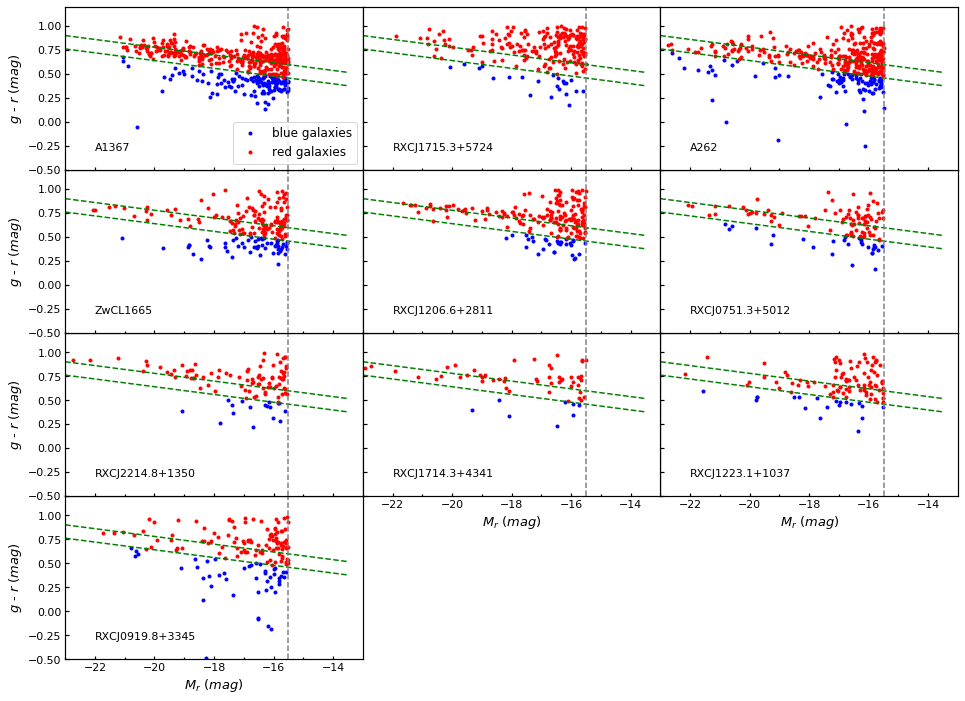}
    \caption{Colour-magnitude diagram for each cluster. Blue point highlights the "blue" galaxies while the red points highlights the "red" galaxies. Green dashed lines indicate the  lines we traced for the red sequence and to do the selection between blue and red galaxies (see Figure \ref{fig:figure10}). Black dotted line marks an absolute magnitude, M$_r$ $=$ -15.5 mag, limit that we used in all the clusters. }
    \label{fig:figure11}
\end{figure*}

Figure \ref{fig:figure11} shows the colour-magnitude diagrams of the individual clusters inside  1$R_{200}$  and up to the absolute magnitude, M$_r$ = -15.5 mag (dashed line on each panel). We overplotted in this figure the red sequence fitted in the global CMD and the line used for the blue and red galaxy separation. Overall,  the  global red sequence line traces well the red sequence of each individual cluster. Note that the few  bright galaxies without available redshifts  are also included in this figure. Interestingly, the number of blue galaxies varies from cluster to cluster, with some of clusters having very few blue galaxies, especially large blue galaxies (for example clusters RXCJ1223.1+1037, RXCJ0751.3+5012, and  RXCJ2214.8+1350) while there are others where the blue galaxy population is larger (see A1367).

\begin{figure}

	\includegraphics[width=\columnwidth, height=17cm]{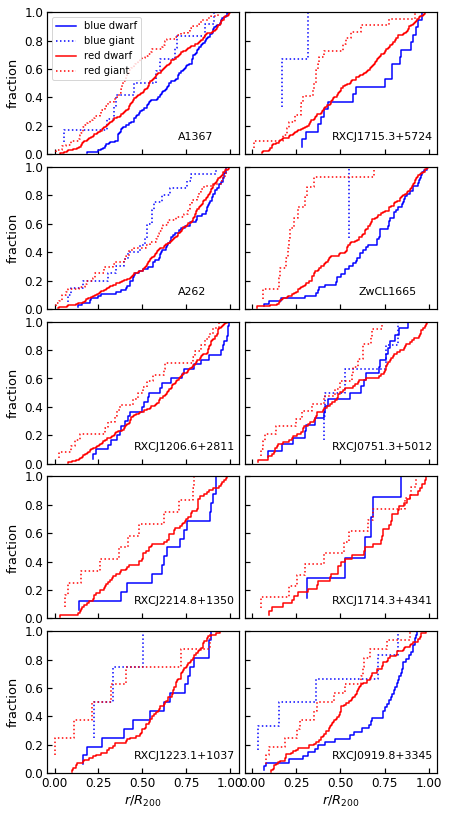}
    \caption{Individual cumulative fractions as a function of the distance  to the cluster centre for the blue and red galaxies. Blue and red dashed lines  correspond to the 'blue' dwarfs and 'red' dwarf respectively, while blue and red filled lines lines correspond  to 'blue' giants and 'red' giants respectively. }
    \label{fig:figure12}
\end{figure}

Figure \ref{fig:figure12} shows the cumulative distribution functions of the blue and red galaxies plotted as a function of their clustercentric distance. We quantified the differences between the blue and red dwarf distributions (filled lines)  by using the KS test (p-values in Table \ref{tab:table5}). According to this test, three clusters (A1367, ZWCL1665 and RXCJ0919+3345)  show blue and red dwarf distributions that are statistically different; in those cases, red dwarfs are located  closer to the cluster centre. We find no correlation with clusters mass, velocity dispersion or dynamical status (see Table \ref{tab:table2}). In contrast, the remaining eight clusters show a similar radial distribution for blue and red dwarfs.

 When checking the cumulative fraction for giant galaxies (dashed lines in Figure \ref{fig:figure12}),  we see that overall  red  giants tend to be distributed  closer to the cluster centre than blue giant galaxies. Note that in some clusters, like RXCJ1206.6+2811 and RXCJ2214.8+1350 there are no blue giant galaxies inside $R_{200}$ so all the bright galaxies in those clusters are red (see also  Figure \ref{fig:figure11}).

\section{discussion}
\label{sec:sec5}

\subsection{Environmental effects}

In the previous sections we studied the spatial distribution of dwarf galaxies, their colours, S\'ersic indices,  and effective radii for each cluster. In the following, we discuss how these properties change depending on the environment;  whether there are differences in the properties of the dwarfs that depend on the mass, velocity dispersion or dynamical status of the cluster. 

\subsubsection{Distribution of dwarfs vs. giants}
In Section 4.1  we showed the distribution of the dwarf candidates in the different clusters of the sample and saw that they dominate in number the galaxy population of the clusters. When comparing the clustercentric distribution of giants and dwarfs, we find that the distribution of giants and dwarfs is different in five of the clusters. In these cases, the bright galaxies are located closer to the cluster centre.  In particular, these cases correspond to the more massive clusters of the sample, as given by $M_{500}$ mass (values that come from X-ray flux measurements, see Fig. \ref{fig:figure7} and Table \ref{tab:table2}). When using the velocity dispersion as a mass ($M_{200}$) indicator the correlation is  less clear.  We quantified the ratio between the fraction of dwarfs inside 0.3$R_{200}$ and the fraction between 0.3$R_{200}$ and 1$R_{200}$ (see the second row in Fig. \ref{fig:figure13}). We found that in all cases this ratio is lower than 1, meaning that the fraction of dwarfs is lower at short distances from the cluster centre, in agreement with earlier results in the literature \citep[e.g.][]{Popesso2005,Rude2020}. A similar finding was noted in \cite{Pavel2018} where a lack of UDGs is reported near the cluster centres. 

The lack of dwarfs in the cluster core  can be explained by several effects. One is tidal disruption  of the dwarfs in the very inner regions  \citep[see][]{Popesso2005}. Another possibility is that giants suffer stronger dynamical friction, making them spiral inward to the cluster centre.  While this mass segregation has been studied theoretically \citep{Kim2020} and observationally \citep{Roberts2015}, there is still no consensus about the findings, as some studies find significant segregation while others find weak segregation in galaxy groups  and clusters. We defer a more detailed analysis of the  distribution of dwarfs and giants to a follow-up study.

\subsubsection{The dwarf fraction in clusters}

\begin{figure}
	\includegraphics[width=\columnwidth]{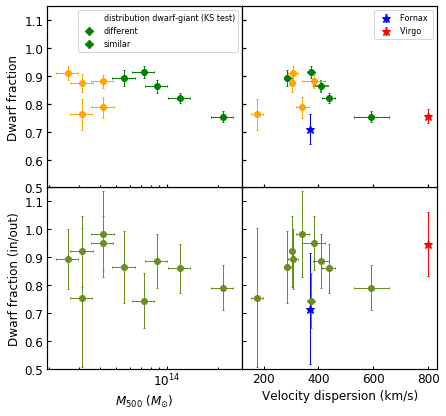}
    \caption{{ Top panels}: Dwarf fraction inside $R_{200}$  as a function of cluster mass $M_{500}$  (left) and cluster velocity dispersion (right). Green colour points indicate whether the distribution of dwarfs and giants is different (based on the KS-test), while orange indicates that both distribution are similar. {Bottom panels}:Ratio of the  dwarf fraction in the inner region and in the outermost regions of the cluster versus the same quantities.}
    \label{fig:figure13}
\end{figure}

We find that the dwarf fraction defined as the ratio between the number of dwarfs over  the total number of galaxies (dwarfs plus giants) within $R_{200}$, in all the clusters of our sample is larger than $\sim$ 0.7. In order to compare these dwarf fractions with well-studied, nearby clusters, such as  Fornax (catalogue from \cite{Venhola2018}, $\sigma_c$ $\sim$374 km/s \citep{Drinkwater2001}) and Virgo  (extended Virgo catalogue from \cite{Kim2014}, $\sigma_c $ $\sim$800 km/s) we estimated the fraction of dwarfs in these clusters by selecting their galaxies in the same way as we did for our   clusters (down to an absolute magnitude of M$_r$ $\leq$-15.5 mag) as is shown in Fig. \ref{fig:figure13} (right panels).  When comparing to Fornax, whose velocity dispersion falls into the regime of most of our clusters, we find that its fraction is smaller than in the clusters (regardless of the region where this fraction is measured) of our sample with as similar mass. It may be that there is still  contamination by background galaxies present in our sample that could affect the dwarf fraction (Section 3.2.1) as most of the bright dwarfs in the Fornax sample are likely confirmed cluster members  \citep{Venhola2018}. Additionally, there seems to be no relation between the dwarf fraction and the dynamical status of the clusters.  Also, there is no clear relation with the distribution of dwarfs and giants (different colours in  Fig.\ref{fig:figure13}).

As regards to  Virgo, that cluster also has a relatively low fraction of dwarfs, however given its high velocity dispersion or mass (M = 1.2 $\times$ $10^{15}$ M$_{\odot}$, \citealt{Fouque2001}), a comparison with the clusters of our sample, which are  less massive, is not simple.

We do not measure a clear decrease  in the dwarf fraction as a function of cluster mass, as seen in \citet{Popesso2005}, who analysed  69 clusters in the $z$-band. Measuring the dwarf-to-giant ratio within a radius of 1 Mpc they found an anti-correlation with the cluster mass, velocity dispersion, and X-ray luminosity, however, the anti-correlation they found is less clear when the dwarf fraction is measured withing $R_{200}$  so in agreement with our result here. Although, their  definition for dwarfs is different.

\subsubsection{S\'ersic indices and Effective Radii across the clusters}
\begin{figure}
	\includegraphics[width=\columnwidth]{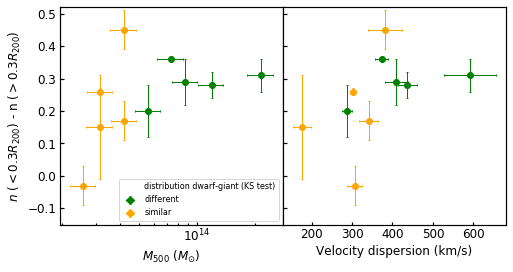}
    \caption{Difference between the  mean S\'ersic index value in the inner regions of the cluster ($r\leq$0.3$R_{200}$) and  the mean value in the outer regions (0.3$R_{200}$ $<$ $r$ $\leq$1$R_{200}$) for the dwarf ellipticals  as a function of cluster mass $M_{500}$  (left) and cluster velocity dispersion (right). Green colour points indicate whether the distribution of dwarfs and giants is different (based on the KS-test), while orange indicates that both distribution are similar.  }
    \label{fig:figure14}
\end{figure}

\begin{figure}
	\includegraphics[width=\columnwidth]{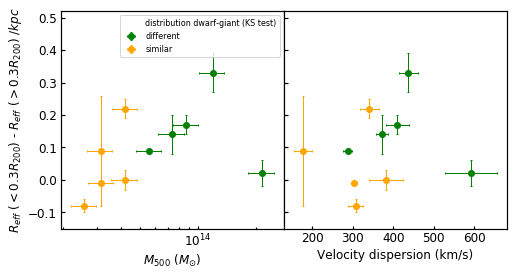}
    \caption{Difference between the mean value of effective radius in the inner regions of the cluster ($r\leq$0.3$R_{200}$) and  the mean value in the outer regions (0.3$R_{200}$ $<$ $r$ $\leq$1$R_{200}$) for the dwarf ellipticals  as a function of cluster mass $M_{500}$  (left) and cluster velocity dispersion (right). Same symbols and colours as in Figure \ref{fig:figure14}. }
    \label{fig:figure15}
\end{figure}

We find hints that dwarfs are more concentrated  (higher S\'ersic index) in the  central regions of the clusters (Fig. \ref{fig:figure14}) and have larger effective radii (Fig. \ref{fig:figure15}). In \citet{Venhola2019}, the  brighter dwarfs of the Fornax cluster (M$_r$ $<$ -16.5 mag) tend to be more concentrated at shorter distances from the cluster centre (i.e. have large S\'ersic indices),  and  they are also found to be smaller at those distances. The argument given in that work is in agreement with tidal interactions and a slow quenching in the inner region of a galaxy, possibly combined with extra central star formation making these galaxies more concentrated. Note that for fainter dwarfs their relation is opposite: such galaxies are larger in the inner parts of the cluster, with similar S\'ersic indices. \citet{Janz2016}  state that dwarfs located in the high-density environment of the Virgo cluster tend to be smaller, a picture that is in agreement with the Fornax dwarf sample. Our results for the S\'ersic indices are compatible with nearby clusters, but for $R_{\rm eff}$ the situation is more complicated. Since the effects are likely to be strongly dependent on the mass of the dwarf, a detailed comparison is needed. We postpone this is to our next paper (Choque Challapa et al. in prep.).

In addition, when comparing  the average value of the S\'ersic index in the inner  and outer region of the clusters as is shown in Fig. \ref{fig:figure14}, there seems to be a difference between the least massive and most massive clusters.  This is, the  average  $n$  value is larger in the inner region (by about 0.3) for almost the full range of  cluster mass or velocity dispersion (see  also Appendix; top panels in Fig. \ref{fig:figure17}) and decreases at further distances from the cluster core (bottom panels in Fig. \ref{fig:figure17});  it tends to be a bit larger for the more massive clusters. Tidal interaction and/or ram-pressure stripping might explain why dwarfs become more concentrated at the cluster core and it seems to be that these effects are stronger in the more massive clusters.  With regards to the difference in spatial distribution between the dwarf and giants galaxies in the cluster, we do not find any clear correlation as a function of the S\'ersic index difference (different colours in Fig. \ref{fig:figure14}).  In a similar way, we do not find a  correlation  between the ratio of mean  S\'ersic indices and the dynamical status. Regarding the effective radius, when comparing its average value in the inner and outer regions, there is a weak tendency for this difference to be larger for more massive clusters. 

\subsubsection{Colours of galaxies}

In the same way that most galaxies in clusters are dwarfs, most of them are red. Additionally, three clusters  of our sample (A1367, ZwCl1665, RXCJ0919+3345) show blue and red dwarf spatial distributions that are statistically different (by taking a KS test, Table \ref{tab:table5} and Fig. \ref{fig:figure12}). In those cases, red, quiescent dwarfs, are located closer to the cluster centre, as has been previously reported on other clusters \citep[e.g.][]{Popesso2005}. We checked whether these three clusters have common properties, such as cluster mass, velocity dispersion or dynamical status, but this is not the case (see Table \ref{tab:table2}). In contrast, the remaining eight clusters show radial distributions that are statistically similar for blue and red dwarfs. For completeness, we also checked the distribution  for giant galaxies (Fig. \ref{fig:figure11}, \ref{fig:figure12}), and in some clusters like RXCJ1206.6+2811 and RXCJ2214.8+1350 there are no blue giant galaxies inside $R_{200}$.

\begin{figure}
	\includegraphics[width=\columnwidth]{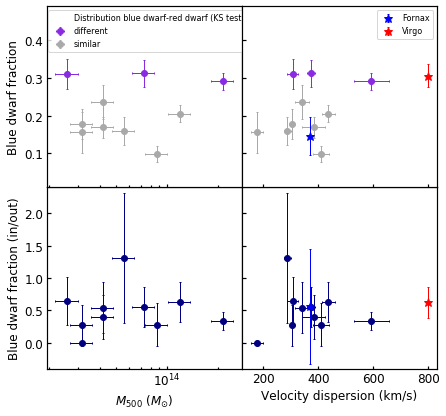}
    \caption{ {Top panels}: Blue dwarf fraction inside  $R_{200}$ as a function of cluster mass $M_{500}$  (left) and cluster velocity dispersion (right). Purple colour points indicate clusters for which the distribution of blue and red dwarfs is different (based on a KS test), while grey indicates that both distribution are similar. {Bottom panels}: Ratio of the blue dwarf fraction in the inner and outer regions of the cluster as a function of cluster mass $M_{500}$  (left) and cluster velocity dispersion (right).}
    \label{fig:figure16}
\end{figure}

When we look for a trend on the the blue dwarf fraction as a function of  the cluster mass and velocity dispersion (top panels in Fig. \ref{fig:figure16}), we find no correlation between these parameters, but we find that the blue dwarf fraction in all the clusters is less than 0.5. Moreover, when we compare with the Fornax and Virgo clusters, the blue dwarf fraction for Fornax is a bit smaller than in most  clusters of our sample, while Virgo has a higher fraction,  similar to three of our clusters. Also,  the clusters with the highest fraction of blue dwarfs are those where the blue dwarfs are distributed differently than red galaxies, such that red dwarfs are generally situated closer to the cluster centre than blue ones.  This agrees with the fact that the ratio of the blue dwarf fraction in the inner and the outer regions of the clusters is in most cases less than 1, indicating that the blue dwarf fraction increases when going away from the centre (see ratio bottom panels in Fig. \ref{fig:figure16}).
We find that these three cases with the highest blue dwarf fraction are not special, as far as their dynamical cluster status is concerned. A possible explanation for these three cluster is that  we are seeing the cluster together with end-on filaments in front or behind, producing a line-of-sight contamination with blue dwarfs, which we expect to be more prevalent in filaments.

\section{Summary and Conclusion}
\label{sec:sec6}

In this work, we analysed a set of twelve nearby clusters from the KIWICS survey, studying the properties of the dwarf population in each cluster for galaxies as faint as a total $r$-band absolute  magnitude, M$_r$ $=$ -15.5 mag. We investigated their distribution in the cluster and their relative fraction, as compared to giant galaxies, and also determined  their S\'ersic index $n$ and their effective radius, $R_{\rm eff}$. Additionally, we studied the colour distribution of dwarfs  to study differences between clusters and correlating them with their physical parameters, to better understand how the environment might affect the dwarf population. Our main results are summarised below,

\begin{itemize}

    \item The dwarf galaxy population dominates by number in all the clusters but their distribution in each cluster is different: they tend to be distributed across the cluster but not always homogeneously. Regarding their distribution compared with giant galaxies, in half of the clusters of our sample, dwarfs are distributed in a statistically different way across the clusters than giants, with giants being closer to the cluster centre. These cases correspond to the more massive clusters of the sample. We find that  the dwarf fraction in each cluster it is greater that $\sim$ 0.75 (Table \ref{tab:table5}),  with small scatter at a fixed mass. Furthermore, when we compare the dwarf fraction in the inner and outer region of the cluster, we find that the dwarf fraction is larger in the outer regions (Fig. \ref{fig:figure13}).

    \item We find that the average S\'ersic index for those dwarfs we classified as early type or quiescent tends to increase at shorter distance from the cluster centre (Table \ref{tab:table6}). This trend is found in all the clusters regardless of their mass. We do not find similar trends for $R_{\rm eff}$.

    \item When exploring the colour distributions, we find that although red dwarf galaxies dominate in all the clusters of our sample, with the total blue fraction always less than $\sim$0.4 (Table \ref{tab:table5}),  the fraction of blue dwarfs is about 50\% lower in the central regions than in the outer parts of the cluster (Fig. \ref{fig:figure16}).

\end{itemize}

 To further understand the properties of dwarf galaxies in the different environments one limiting factor has been the absence of redshifts for the fainter dwarfs, to be able to determine their cluster membership.  Although SDSS is providing many redshifts, better and deeper surveys are needed to provide enough objects to study properties like substructure in the dwarf population, which can tell us about how the cluster environment is influencing the formation and evolution of dwarf galaxies.
This issue might be improved in the near future when the WEAVE nearby spectroscopic cluster survey,  that includes all the  clusters of our photometric survey, will provide us with thousands of new redshifts per cluster. In this way, our background contamination will drop and  our analysis will be more accurate. While WEAVE will obtain intermediate resolution spectra in its MOS mode for all dwarf galaxies in the central regions down to a magnitude M$_r<$-15 mag, the mIFU mode  will allow us  to obtain 2-dimensional internal properties  of a selected subsample of dwarfs, giving us further information about  the kinematic, mass distribution, stellar populations and metallicities for the dwarf population in these nearby clusters \citep[e.g.][]{scott2020}, which can then  be compared with the properties of the clusters.

\section*{Acknowledgements}
We thank the anonymous referee for their comments that have improved this work. RP and NCC acknowledge financial support from the European Union’s Horizon 2020 research and innovation programme under Marie Sk\l{}odowska-Curie grant agreement no. 721463 to the SUNDIAL ITN network. NCC acknowledges the financial support from CONICYT PFCHA/DOCTORADO BECAS CHILE/2016 - 72170347. PEMP is supported by the Netherlands Research School for Astronomy (Nederlandse Onderzoekschool voor Astronomie, NOVA), Project 10.1.5.6.  We also thank to Julia Healy for providing the spectroscopic  data for one of our analysed objects (cluster A262).

This work is based on several  observations made with the Isaac Newton Telescope, therefore, we thanks to all the people that contributed doing the observations of the KIWICS survey.

\section*{Data Availability}

The data underlying this article will be shared upon a request to the corresponding authors. The raw data is  stored in the Isaac Newton Group Archive. 



\bibliographystyle{mnras}
\bibliography{references} 




\appendix

\appendix

\section{SExtractor configuration}
\label{sec:appA}

Main SExtractor  configuration parameters used for the smaller object detection.
\newline

DETECT MINAREA = 20, DETECT THRESHOLD = 2, ANALYSIS THRESHOLD = 2, BACK SIZE = 64.
\newline

For very large objects not detected with the above parameters, 
\newline

DETECT MINAREA = 10000, DETECT THRESHOLD = 50, ANALYSIS THRESHOLD = 50, BACK SIZE = 21000.

\section{S\'ersic indices and Effective Radii}

In this appendix, we report the plots showing the average value  of the S\'ersic index and effective radius  inside a distance of 0.3$R_{200}$ and outside (0.3$R_{200}$ $<$ $r$ $<$ 1$R_{200}$) for the different clusters as a function of their mass ($M_{500}$) and velocity dispersion. 

\begin{figure}
	\includegraphics[width=\columnwidth]{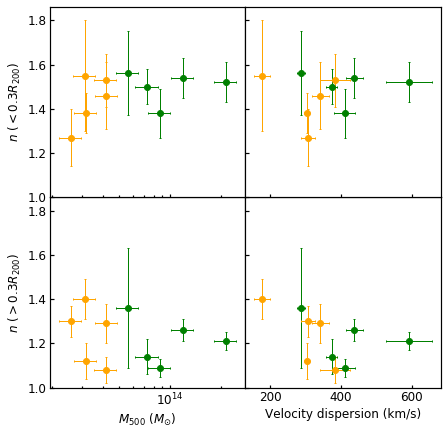}
    \caption{S\'ersic index for dwarf ellipticals as a function of velocity dispersion and cluster mass, $M_{500}$ in the inner (top panels) and outer (bottom panels) region of each cluster. Green colour points indicate whether the dwarf and giant distributions are statistically different (based on the KS test), while orange indicates both distribution are statistically similar.}
    \label{fig:figure17}
\end{figure}

\begin{figure}
	\includegraphics[width=\columnwidth]{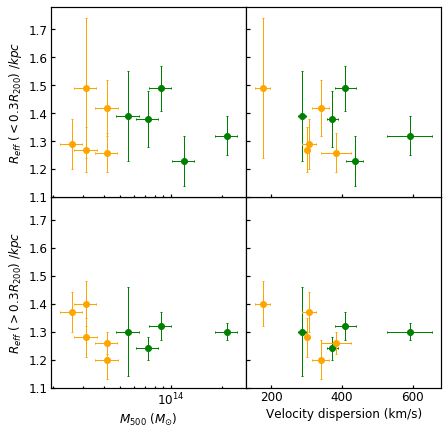}
    \caption{Effective radius for dwarf ellipticals as a function of velocity dispersion and cluster mass, $M_{500}$ in the inner (top panels) and outer (bottom panels) region of each cluster. Green colour points indicates whether the dwarf and giant distributions are statistically different (based on the KS test), while orange indicates both distribution are statistically similar.}
    \label{fig:figure18}
\end{figure}

\bsp	
\label{lastpage}
\end{document}